\documentclass[conference]{IEEEtran}

\usepackage[dvips]{graphicx}
\usepackage[cmex10]{amsmath}
\usepackage{array}
\usepackage{mdwmath}
\usepackage{mdwtab}
\usepackage{booktabs}
\usepackage{multirow}
\usepackage{CJK}
\usepackage{subfig}
\usepackage{multirow}
\usepackage{float}
\usepackage{enumerate}
\usepackage{algorithm}
\usepackage{algpseudocode}
\usepackage{diagbox }
\newcommand{\tabincell}[2]{\begin{tabular}{@{}#1@{}}#2\end{tabular}}

\makeatletter

\newcommand{\Rmnum}[1]{\expandafter\@slowromancap\romannumeral #1@}
\makeatother
\ifCLASSINFOpdf
  \else
\fi

\begin{document}

\title{On the randomness analysis of link quality prediction: limitations and benefits}

\author{
\IEEEauthorblockN{Shi Xiaofei }
\IEEEauthorblockA{Macau University of Science and Technology\\
Macau, P.R.China\\
Email: 604767519@qq.com}

\and

\IEEEauthorblockN{Liao Wenxing }
\IEEEauthorblockA{Information Engineering College\\
Shaoguan University \\
Shaoguan 512000, P.R.China\\
Email: liaowenxing@sgu.edu.cn}

}

\maketitle

\begin{abstract}
In wireless multi-hop networks, such as wireless sensor networks, link quality (LQ) is one of the most important metrics and is widely used in higher-layer applications such as routing protocols. An accurate link quality prediction may greatly help to improve the performance of wireless multi-hop networks. Researchers have proposed a lot of link quality prediction models in recent years. However, due to the dynamic and stochastic nature of wireless transmission, the performance of link quality prediction remains challenging. In this article, we mainly analyze the influence of stochastic nature of wireless transmission on link quality prediction model and discuss the benefits in the application of wireless multi-hop networks with the performance-limited link quality prediction models.

\end{abstract}

\section{Introduction}
In wireless multi-hop networks, the performance of higher-layer applications, such as routing, power control and energy consumption [1,2], relies heavily on the high accuracy of LQ prediction. Routing protocols for example, if the LQ is obtained before making routing decision, the weak connections are eliminated and the stability and robustness are improved. Therefore, the LQ prediction between nodes has great significance in wireless multi-hop networks. 

However, in most wireless networks, the LQ between a transmitter and a receiver is hard to accurately predict. Radio wave propagation experiences reflection, blocking, scattering and diffraction. These factors have an adverse effect on the transmitted signal that travels in a wireless channel [3]. Commonly, the received wireless signal consists of two parts, one of which is distance-dependent path loss and the other is small-scale fading [4]. Because of the randomness of small-scale fading, the received signal level is unpredictable. The log-normal shadow fading channel model [5] for example, the stochastic fading component is chosen from a normal probability density function. Meanwhile, the mobility of nodes in wireless multi-hop networks brings additional randomness in received signal level. 

Generally, in LQ prediction models, there are mainly two types of LQ metrics[4], the physical metrics, such as Received Signal Strength Indicator (RSSI) and logical metrics, such as Packet Reception Ratio (PRR). Unfortunately, both types of metrics are seriously affected by the randomness of the received signal and show a strong unpredictability at the receiving nodes. Although there is a strong link between the transmitter and receiver, accidental packet loss may still occur. 

At present, the researchers have proposed a number of LQ prediction methods, including traditional optimization algorithms, such as (EWMA) [6,7] and Machine learning-based algorithms, such as (fuzzy and svr)[8,9]. The historical values of LQ metrics such as Signal-to-Noise Ratio (SNR), Received Signal Strength Indicator (RSSI), Link Quality Indicator (LQI), Frame Error Rate (FER), Pack Lost Rate (PLR) and Packet Delivery Rate (PDR) are used as the input parameters of prediction models [4]. The output parameters mainly include the LQ over a period of time, such as the package reception or the received signal strength for the next broadcast cycle.  

The mainstream LQ prediction models are generally machine learning-based, and can be divided into two types, the offline [3,10,11] and online [12-16] learning models. In [11] and [12], the authors propose an online learning model (TALENT) and an offline learning model (4C) respectively. Both LQ prediction models focus on the improvement of packet delivery rate. The authors use the PRR,RSSI,SNR and LQI as the input parameters to train the models. 

Consider the network model, most of the LQ prediction models are proposed for the static or quasi-static networks, such as MESH and WSN [12-14,17-19]. While for the networks with mobility, (MANET for example) [3,11,20], the additional network dynamic information is necessary for LQ prediction models. Otherwise, authors in [17,19] propose LQ prediction models for indoor and outdoor wireless sensor networks respectively. 

Although a large number of LQ prediction models have been proposed for different application scenarios and different network models, most of these prediction models adopt similar research idea. The accurate prediction of LQ relies on the time and space correlation of wireless links. That is the LQ has certain regularity in space and time. However, as discussed above, due to the randomness of wireless transmission and the mobility of network nodes, the overall prediction accuracy is not high, and most studies have not conducted in-depth analysis on this issue. 

In this paper, we propose a machine learning-based LQ prediction model. In the proposed model, we apply the historical values of both the physical and logical metrics as input parameters and the LQ at the next moment as the output parameter. By deeply analyzing the performance of the proposed model, we discuss the impact of stochastic nature of wireless transmission on LQ prediction applications. 

The main contributions of this paper are as follows.

\begin{enumerate}
\item  We define the randomness metrics of both the prediction model and the wireless link, which is helpful for in-depth understanding of the impact of the stochastic nature of wireless links.

\item  We analyze the performance bottleneck of LQ prediction model and discuss the limitation brought by the randomness of wireless transmission. 

\item  We point out the application direction of performance-constrained LQ prediction models in wireless multi-hop networks.  
\end{enumerate}

The rest of this paper is organized as follows. In section \Rmnum{2}, we design a machine learning-based wireless LQ prediction model and analyze the randomness of wireless transmission in detail. In section \Rmnum{3}, we analyze the performance of the proposed LQ prediction model with different values of randomness metric. Then we mainly analyze the limitations of the prediction model from two aspects: the randomness of the prediction output and the accuracy of the prediction. In section \Rmnum{4}, the application of the performance-limited LQ prediction model is discussed. Finally, we conclude this paper in section \Rmnum{5}.


\section{Models}

\subsection{Link Quality Prediction model}
In this paper, we consider a wireless sensor network with random deployment of sensor nodes. In such network, the sensor nodes apply the LQ prediction models to get the quality of their wireless links in order to optimize routing, avoid disconnection, and make other network optimization.

The nodes periodically broadcast HELLO packets to their neighbors and receive HELLO packets from their neighbors. In this way, the sensor nodes are easy to obtain historical values of LQ metrics. The proposed model is as shown in Fig 1. 

\begin{figure}[h]
\centering
\vspace{0.1cm}
\includegraphics[width=3.2 in]{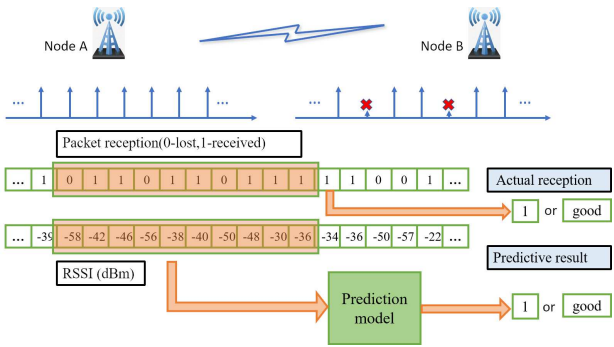}
\centering{\caption{The structure of wireless LQ prediction model}\label{}}
\end{figure}

Suppose there are two nodes A and B, node A broadcasts HELLO packets periodically and node B receives these HELLO packets. Because of the fading characteristics of wireless wave propagation, HELLO packets may lost at node B and the reception of HELLO packets becomes random. In the proposed LQ prediction model, we use the historical information of HELLO packets for the previous K broadcast cycles as the input parameters to train the model. The historical information is the combination of both the logical and physical LQ metrics, the packet reception and the corresponding RSSI. 

In terms of prediction outputs (the labels), we mainly analyze two options. One is the HELLO packet reception for the next broadcast cycle, which is a two-classification problem, and the other is the LQ level, which shows in Good(G), Medium Good(MG), Medium Bad(MB) and Bad(B), for the next period. Definitions of these labels are shown in TABLE \Rmnum{1} and TABLE \Rmnum{2}. 

\begin{table}[!t]
\renewcommand{\arraystretch}{1.3}
\caption{Labels in two-classification case}
\label{table_example}
\centering
\begin{tabular}{|c|c|}
\hline
\textbf{label}  & \textbf{definition} \\
\hline
Received & \tabincell{c}{The HELLO packet is successfully received \\ at the next broadcast cycle} \\
\hline
Lost &\tabincell{c} {The HELLO packet is lost at \\ the next broadcast cycle} \\
\hline
\end{tabular}
\end{table}

\begin{table}[!t]
\renewcommand{\arraystretch}{1.3}
\caption{Labels in four-classification case}
\label{table_example}
\centering
\begin{tabular}{|c|c|}
\hline
\textbf{label}  & \textbf{definition} \\
\hline
G & \tabincell{c}{Receive 3 HELLO packets during \\the following 3 broadcast cycles} \\
\hline
MG & \tabincell{c}{Receive 2 HELLO packets during \\the following 3 broadcast cycles}\\
\hline
MB & \tabincell{c}{Receive 1 HELLO packets during \\the following 3 broadcast cycles}\\
\hline
B & \tabincell{c}{Receive 0 HELLO packets during \\the following 3 broadcast cycles}\\
\hline
\end{tabular}
\end{table}

For a more comprehensive analysis, we apply different machine learning models, such as Neural Networks(NN), Random Forest(RF), Decision Tree(DT),XGBoost and Gradient Boosting Decision Tree(GBDT). Then we compare the performance of these models and choose a best model for in-depth analysis.

\subsection{Link model}
In this paper, we introduce a log-normal shadow fading model to describe the wireless link between two sensor nodes. Let $s(u,v)$ denotes the distance between node $u$ and node $v$. The transmitting power of node $u$ and the receiving power of node $v$ are denoted as $p_t(u)$ and $p_r(v)$ respectively. The signal attenuation is given by
\begin{equation}
\beta(u, v)=10 \log _{10}\left(\frac{p_{t}(u)}{p_{r}(v)}\right) dB
\end{equation}

According to the log-normal shadow fading model, the signal attenuation comprises two components: 
\begin{equation}
\beta(u, v) = \beta_{1}(u, v)+\beta_{2}
\end{equation}

$\beta _1(u, v)$ is deterministic geometric component and it is given by 
\begin{equation}
\beta_{1}(u, v)=\alpha 10 \log _{10}\left(\frac{s(u, v)}{1 m}\right) d B
\end{equation}
with the path-loss exponent $\alpha$.

$\beta _2 $  is stochastic component and it is chosen from a normal probability density function.
\begin{equation}
f_{\beta_{2}}\left(\beta_{2}\right)=\frac{1}{\sqrt{2 \pi} \sigma} \exp \left(-\frac{\beta_{2}^{2}}{2 \sigma^{2}}\right)
\end{equation}
with the standard deviation $\sigma $

For given $p_t$ and $p_{r,th}$, we have the threshold attenuation
\begin{equation}
\beta_{th}=10 \log _{10}\left(\frac{p_{t}}{p_{r,th}}\right) d B
\end{equation}

When the signal attenuation $\beta $ is less than the threshold attenuation, the receiver successfully receives the packet from the transmitter. With the increase of distance between the neighbor nodes, the probability of successful reception decreases and this reflects the actual wireless environment. From [5, eq. (13)], we get the packet delivery rate of two nodes with distance $d$.
\begin{equation}
p(d)=\frac{1}{2}-\frac{1}{2} e r f\left(\frac{10 \alpha}{\sqrt{2} \sigma} \log _{10} \frac{d}{r_{0}} dB\right)
\end{equation}
where $r_0$ is the distance when the point-to-point packet delivery rate drops to 50\%.
\begin{equation}
r_{0}=10^{\frac{\beta_{t h}}{10 \alpha}}
\end{equation}

The curve of $p(d)$ is shown in Fig 2.
\begin{figure}[h]
\centering
\vspace{0.1cm}
\includegraphics[width=3.2 in]{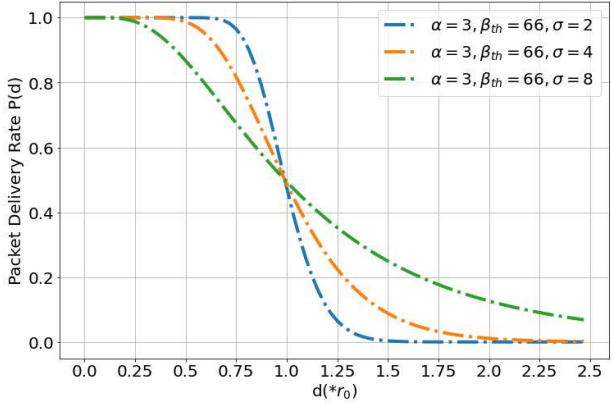}
\centering{\caption{The packet delivery rate of two nodes with distance $d$}\label{}}
\end{figure}

When the distance is relatively small, the packet delivery rate basically remains above 90\%. When the distance is close to $r_0$, $p(d)$ drops sharply which means that the link becomes unstable. When the distance continues to increase, the link becomes stable again and remains disconnected. Therefore, as shown in Fig 2, according to the distance between the neighbor nodes, the wireless channel can be roughly divided into three areas. Obviously, in area around $r_0$, the link is the most unstable and it is an important factor affecting the stability of the wireless multi-hop networks.

\subsection{Randomness metric}
The randomness of wireless transmission is mainly reflected in two aspects: time and space. Consider the common-used link quality metric, the RSSI, we have
\begin{equation}
\begin{aligned}
\beta&=10 \log _{10}\left(\frac{p_{t}}{1mW}\right)-10 \log _{10}\left(\frac{p_{r}}{1mW}\right)\\
&=p_{t}(dBm)-RSSI(dBm)
\end{aligned}
\end{equation}

Let the transmitting power $p_t=50mW$, as shown in Fig 3, with the increase of distance, the RSSI shows a decreasing trend and strong randomness. 
\begin{figure}[h]
\centering
\vspace{0.1cm}
\includegraphics[width=3.2 in]{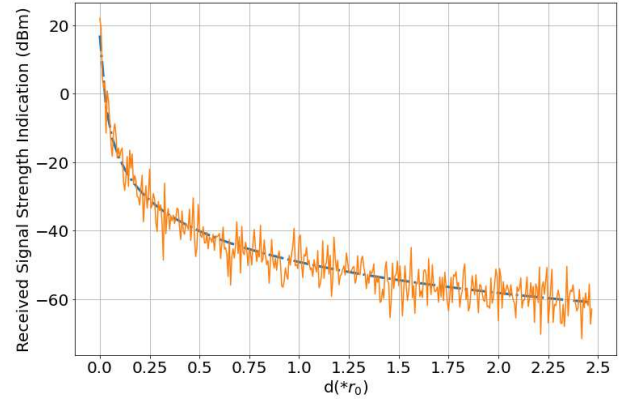}
\centering{\caption{The change of the RSSI over distance}\label{}}
\end{figure}

For two nodes with a given distance, $r_0$ for example, the change of the RSSI over time is as shown in Fig 4. 
\begin{figure}[h]
\centering
\vspace{0.1cm}
\includegraphics[width=3.2 in]{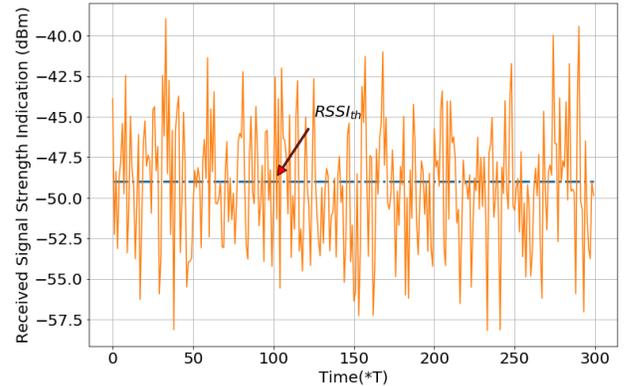}
\centering{\caption{The change of the RSSI over time}\label{}}
\end{figure}

If we set a threshold of RSSI, the randomness of packet reception is as shown in Fig 5.
\begin{figure}[h]
\centering
\vspace{0.1cm}
\includegraphics[width=3.2 in]{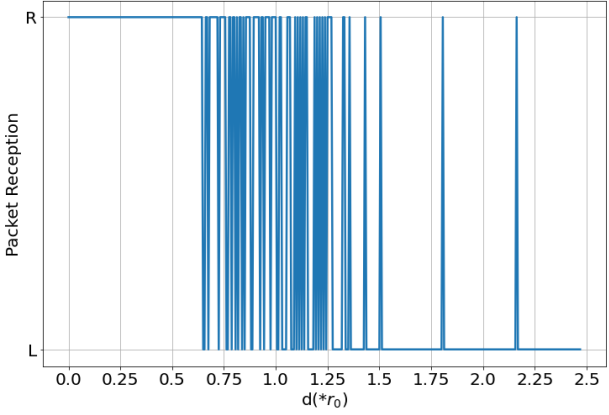}
\centering{\caption{The packet reception over distance with a threshold of RSSI}\label{}}
\end{figure}

From Fig 5, we can see that if the distance between two nodes is around $r_0$, the packet reception shows strong randomness.
 
In our proposed LQ prediction model, we use the historical information of HELLO packets as input parameters and use the packet reception for the next broadcast cycle or the LQ level for the next period as the output label. After we get the historical information (the data part of a sample), because of the randomness of wireless transmission, the label of this sample also shows a certain degree of randomness. As shown in Fig 6, the link condition is good for the pass several broadcast cycles, the LQ level label for this sample may also be marked as bad with a certain probability. 
\begin{figure}[h]
\centering
\vspace{0.1cm}
\includegraphics[width=3.2 in]{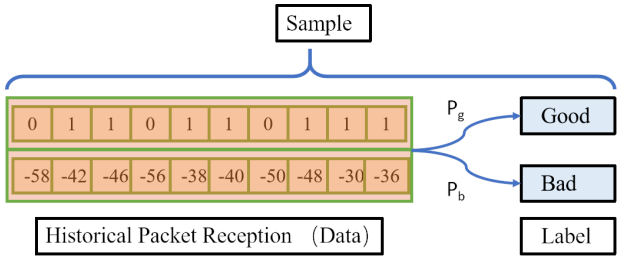}
\centering{\caption{The mislabeling of a given sample because of the randomness}\label{}}
\end{figure}

In order to measure the randomness of a sample set, we define a randomness metric $U$. In the following sections, we analyze the relationship between the performance of the LQ prediction model and the randomness metric $U$. Suppose there are $M$ different labels in the LQ prediction model, in order to train this model, we have the sample set $S$, which is given by
\begin{equation}
S=\left\{\left\langle D_{0}, t_{0}\right\rangle,\left\langle D_{1}, t_{1}\right\rangle, \cdots,\left\langle D_{i}, t_{i}\right\rangle, \cdots\left\langle D_{N}, t_{N}\right\rangle\right\}=\bigcup_{i} S_{i}
\end{equation}
where $N$ is the number of samples in $S$, $D_i$ is the data part of the $i_{th}$ sample, $t_i$ is the label of the $i_{th}$ sample and $S_i$ is the subset of $S$. Consider that the randomness of wireless links in different environments is different, the samples in $S_i$ are obtained from the nodes in the same environment and the subsets satisfy $S_{i} \cap S_{j}=\Phi(i \neq j)$. We define the label set $T$ as
\begin{equation}
T=\{0,1,2, \cdots, M-1\}
\end{equation}

Then, we have a sample $s=\langle D, t\rangle \in S, t \in T$. The randomness metric $U$ of sample set $S$ is defined as
\begin{equation}
U(S)=\sum_{i} \sum_{t \in T} U_{i}(t) R_{i}(t)
\end{equation}
where $R_i(t)$ is the ratio of samples with label $t$ in $S_i$.
\begin{equation}
R_{i}(t)=\frac{\operatorname{size}\left(\left\{\langle D, t\rangle \mid\langle D, t\rangle \in S_{i}, t \in T\right\}\right)}{\operatorname{size}(S)}
\end{equation}

$U_i(t)$ is the randomness matric of a given sample with label $t$ in $S_i$.
\begin{equation}
U_{i}(t)=\sum_{u \in T}-p_{t u} \log _{2}^{p_{t u}}
\end{equation}
where $p_{tu}$ is the probability that label $t$ is mistakenly marked as label $u$. In LQ prediction application, for two nodes with distance $d$, the distribution of  $p_{tu}$ for the samples obtain from these two nodes are shown in TABLE \Rmnum{3} and TABLE \Rmnum{4}.

\begin{table}[!t]
\renewcommand{\arraystretch}{1.3}
\caption{Distribution of $p_{tu}$:Two-classification case}
\label{table_example}
\centering
\begin{tabular}{|c|c|c|}
\hline
\diagbox{t}{u} & Lost & Received \\
\hline
Lost & $1-p(d)$ & $p(d)$\\
\hline
Received & $1-p(d)$ & $p(d)$\\
\hline
\end{tabular}
\end{table}

\begin{table}[!t]
\renewcommand{\arraystretch}{1.3}
\caption{Distribution of $p_{tu}$:Four-classification case}
\label{table_example}
\centering
\begin{tabular}{|c|c|c|c|c|}
\hline
\diagbox{t}{u} & Good & Medium Good & Medium Bad & Bad\\
\hline
G & $P(d)^{3}$ & $3(1-P(d)) P(d)^{2}$ & $3(1-P(d))^{2} P(d)$ &$(1-P(d))^{3}$\\
\hline
MG & $P(d)^{3}$ & $3(1-P(d)) P(d)^{2}$ & $3(1-P(d))^{2} P(d)$ &$(1-P(d))^{3}$\\
\hline
MB & $P(d)^{3}$ & $3(1-P(d)) P(d)^{2}$ & $3(1-P(d))^{2} P(d)$ &$(1-P(d))^{3}$\\
\hline
B & $P(d)^{3}$ & $3(1-P(d)) P(d)^{2}$ & $3(1-P(d))^{2} P(d)$ &$(1-P(d))^{3}$\\
\hline
\end{tabular}
\end{table}
Then, we define the accurate of the sample set as 
\begin{equation}
A(S)=\sum_{i} \sum_{t \in T} R_{i}(t) p_{t t}
\end{equation}

The randomness metric $U$ of sample sets obtain from two nodes with different distance $d$ is as shown in Fig 7.
\begin{figure}[h]
\centering
\vspace{0.1cm}
\includegraphics[width=3.2 in]{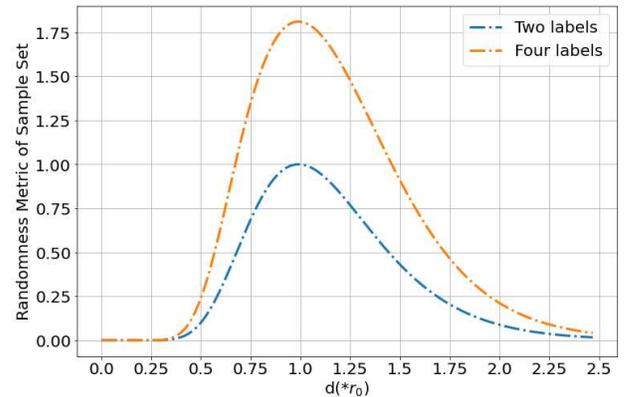}
\centering{\caption{The change of the $U$ over distance:Two-classification and Four-classification cases}\label{}}
\end{figure}

From Fig 7, we can see that when distance is around $r_0$, $U$ reaches the maximum value, which is consistent with the actual link condition. The randomness metric $U$ can well describe the randomness of a sample set.

\section{Performance analysis}
In this section, we mainly analyze the performance of the proposed LQ prediction model. 

Before the analysis, we define the randomness metric of LQ prediction model $U_p$. Let $S_p$ denote the output sample set of the LQ prediction model when the input sample set is $S$. Similar to the definition of $U$, $U_p$ is given by
\begin{equation}
U_{p}(S)=\sum_{t \in T} \sum_{u \in T}-p_{t u}^{\prime} \log _{2}^{p_{t i}^{\prime}} R_{p}(t)
\end{equation}
where $p_{t u}^{\prime}=p(output=u|input=t)$, which denotes the conditional probability that the input label is $u$ when the prediction output label is $t$. 

Considering the uneven sample set, the output of the prediction model may be the same so as to obtain a high accuracy. For example, if 90\% of the samples in the sample set have label $t$, the prediction output of all samples may have label $t$, which can achieve a high accuracy of at least 90\%. Therefore, the maximum accuracy for reference is given by
\begin{equation}
Acc_{\max }=\max \left(A(S), \sum_{i} \max _{t \in T}\left(R_{i}(t)\right)\right)
\end{equation}

In order to train the proposed model, we obtain the samples from two nodes with different distance, which is evenly distributed in (0,$2.5r_0$). 70\% of the samples are used as training sample set $T_r$ and 30\% of the samples are used as test sample set $T_e$. The performance of LQ prediction models with different machine learning algorithms are as shown in TABLE \Rmnum{5} and TABLE \Rmnum{6}.

\begin{table}[!t]
\renewcommand{\arraystretch}{1.3}
\caption{Performance of LQ Prediction Models:Two-classification case}
\label{table_example}
\centering
\begin{tabular}{|c||c||c||c||c|}
\hline
   & \textbf{$ACC$} & \textbf{$Acc_{max}$} & \textbf{$U_p(T_e)$} & \textbf{$U(T_e)$} \\
\hline
Neural Networks &0.8921&0.8985 & 0.4945 & 0.3404 \\
\hline
Random Forest &0.8908	&0.8985&0.4966&0.3404 \\
\hline
Decision Trees &0.8413&0.8985&0.6225&0.3404\\
\hline
GBDT & 0.8913&0.8985&0.4957&0.3404\\
\hline
XGboost &0.8901&0.8985&0.4978&0.3404\\
\hline
\end{tabular}
\end{table}

\begin{table}[!t]
\renewcommand{\arraystretch}{1.3}
\caption{Performance of LQ Prediction Models:Four-classification case}
\label{table_example}
\centering
\begin{tabular}{|c||c||c||c||c|}
\hline
   & \textbf{$ACC$} & \textbf{$Acc_{max}$} & \textbf{$U_p(T_e)$} & \textbf{$U(T_e)$} \\
\hline
Neural Networks &0.7847	&0.7984	&0.7384	&0.6736\\
\hline
Random Forest &0.7812	&0.7984	&0.7391	&0.6736\\
\hline
Decision Trees &0.7097	&0.7984	&1.07	&0.6736\\
\hline
GBDT & 0.787	&0.7984	&0.7497	&0.6736\\
\hline
XGboost &0.7846	&0.7984	&0.7291	&0.6736\\
\hline
\end{tabular}
\end{table}

From TABLE \Rmnum{5} and TABLE \Rmnum{6}, we can see that the accuracy of each LQ prediction model is about the same. In two-classification case, the NN model achieves the highest accuracy while in the four-classification case, the GBDT model achieves the highest accuracy. Nevertheless, the accuracy of the model is less than the reference maximum accuracy $Acc_{max}$. In terms of randomness, the randomness metric of LQ prediction model $U_p$ is slightly larger than the randomness of sample set $U$, which means that the prediction model brings additional randomness while making LQ prediction and the prediction output is still full of randomness. Therefore, in the applications using LQ prediction, affecting by the randomness of wireless transmission, high accuracy is not a necessary condition for the application. In the following analysis, we apply the NN and GBDT models for the two-classification and four-classification cases respectively. 

\subsection{Performance in static randomness environment}

Considering the static wireless multi-hop networks, such as wireless sensor network and wireless mesh network, we obtain the samples from two nodes with the same distance, $d=r_0$ for example. Then we train and test the LQ prediction model using these samples. The performance of LQ prediction models are as shown in Fig 8-11.

\begin{figure}[h]
\centering
\vspace{0.1cm}
\includegraphics[width=3.2 in]{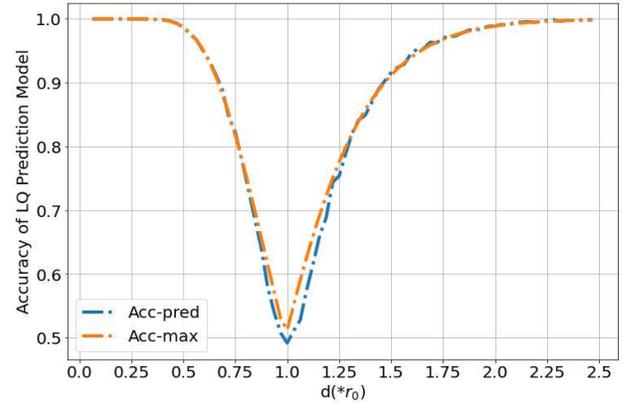}
\centering{\caption{The accuracy of LQ prediction model:Two-classification case}\label{}}
\end{figure}

\begin{figure}[h]
\centering
\vspace{0.1cm}
\includegraphics[width=3.2 in]{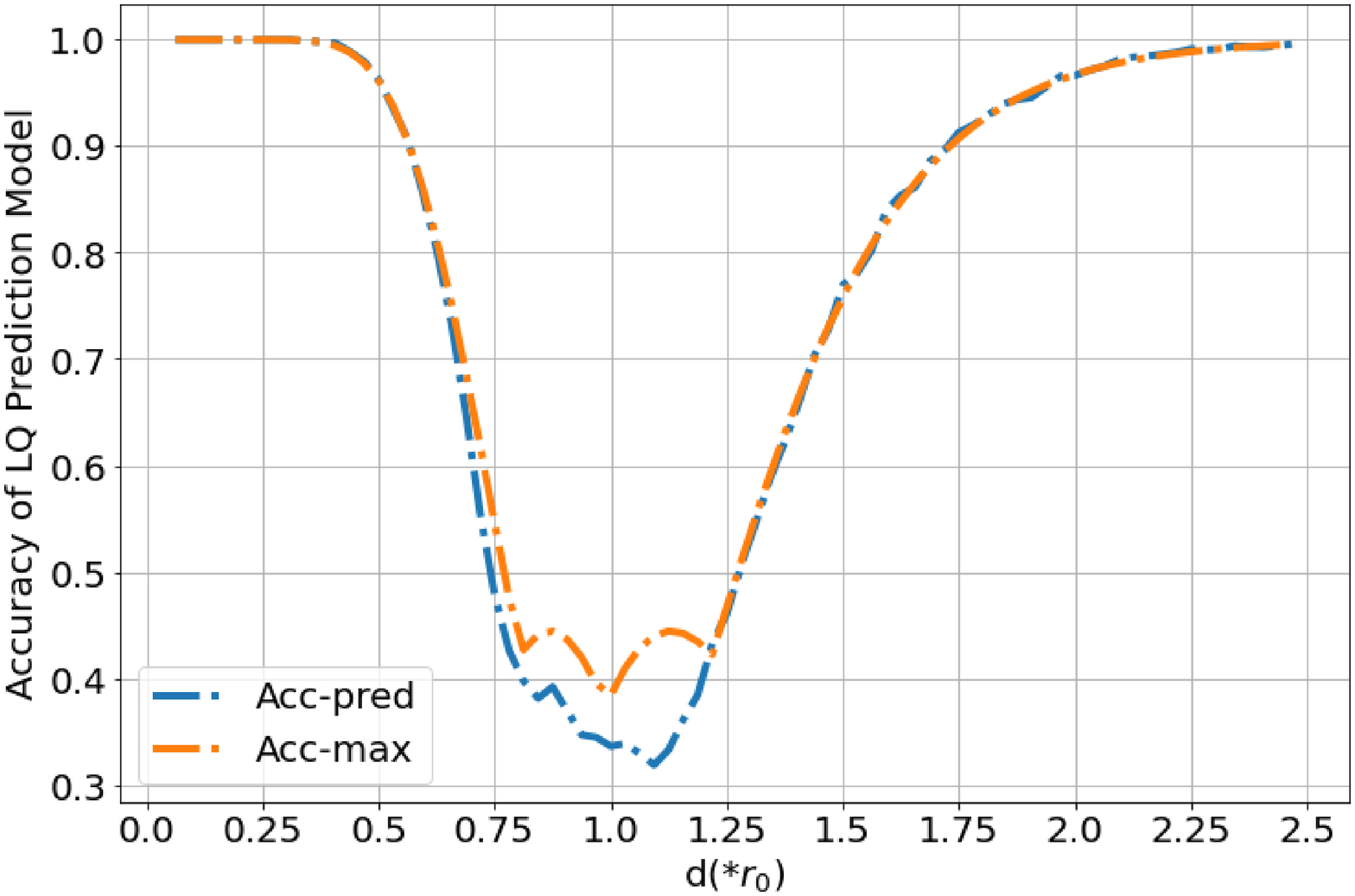}
\centering{\caption{The accuracy of LQ prediction model:Four-classification case}\label{}}
\end{figure}

\begin{figure}[h]
\centering
\vspace{0.1cm}
\includegraphics[width=3.2 in]{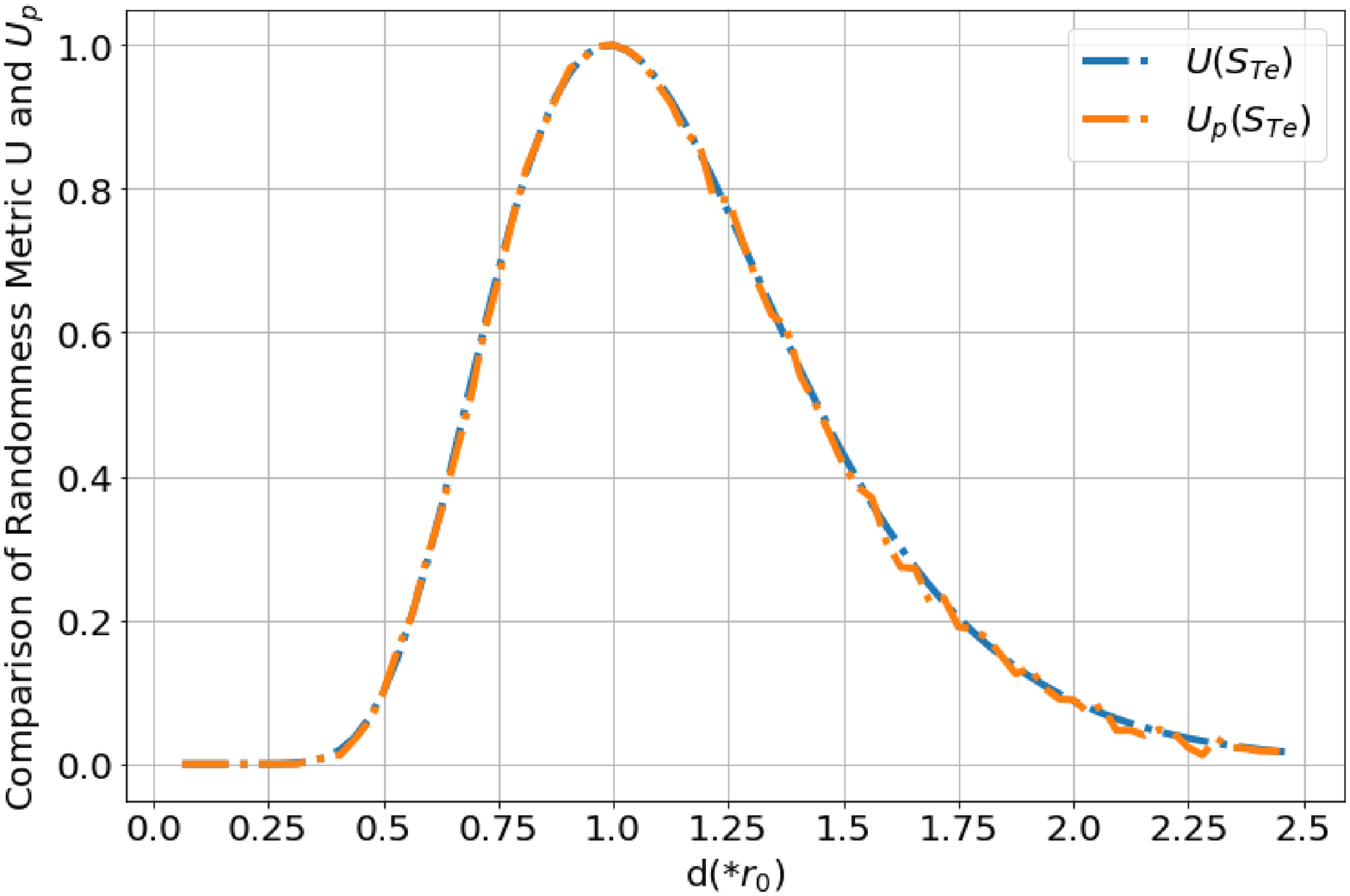}
\centering{\caption{The randomness metric of LQ prediction model:Two-classification case}\label{}}
\end{figure}

\begin{figure}[h]
\centering
\vspace{0.1cm}
\includegraphics[width=3.2 in]{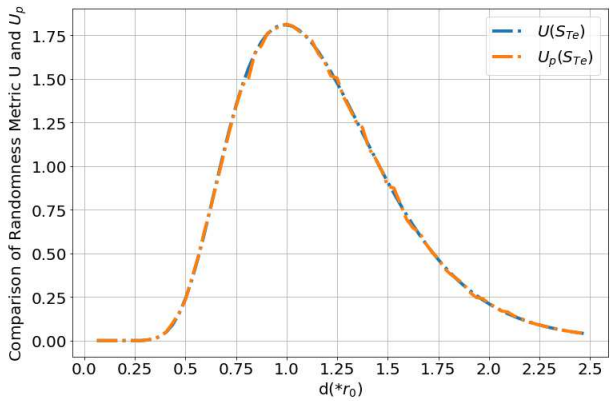}
\centering{\caption{The randomness metric of LQ prediction model:Four-classification case}\label{}}
\end{figure}

Fig 8-9 are the comparison of the accuracy of prediction model and the reference maximum accuracy for two-classification and four-classification cases. Fig 10-11 are the comparison of randomness metric $U_p$ and $U$ for two-classification and four-classification cases. From the figures we can see that with the increase of randomness metric of sample set, the accuracy of the prediction model drops. The prediction output has similar randomness metric as the sample set. When two nodes are very close or far apart, the accuracy of the prediction model is basically equal to the maximum accuracy. This is because when two nodes are very close or far apart, the randomness of wireless transmission is weak and the connection (or disconnection) is strong. As a result, the prediction model is easy to make accurate predictions.

However, when the distance between two nodes is getting close to $r_0$, the randomness of samples increase sharply, as shown in Fig 10 and Fig 11. Even the best prediction model is applied, the prediction output still contains strong randomness and the output of the predictor is basically purely random. As shown in Fig 8 and Fig 9, the accuracy of prediction model has an unacceptably low accuracy when the distance is around $r_0$. Actually, in most wireless multi-hop networks, the link quality is almost unpredictable when the distance is around $r_0$. As a result, when such links are selected as relay links in wireless multi-hop networks, networks become unstable.

\subsection{Performance in dynamic randomness environment}

To further analyze the impact of the randomness on the performance of prediction model, we modify the value of randomness metric $U$ by adjusting the composition of the sample set. Then we train and test the LQ prediction model using the sample sets with different randomness. In this paper, we apply the samples obtained from two nodes with distance dynamically distributed in ($0.5r_0$,$1.5r_0$). The performance of LQ prediction models are as shown in Fig 12-15.

\begin{figure}[h]
\centering
\vspace{0.1cm}
\includegraphics[width=3.2 in]{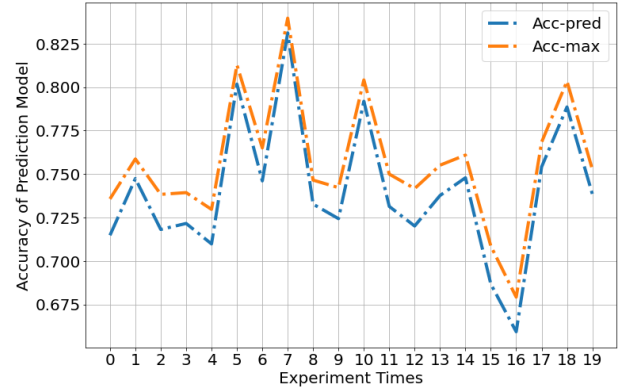}
\centering{\caption{The accuracy of LQ prediction model with randomness selected training sets:Two-classification case}\label{}}
\end{figure}

\begin{figure}[h]
\centering
\vspace{0.1cm}
\includegraphics[width=3.2 in]{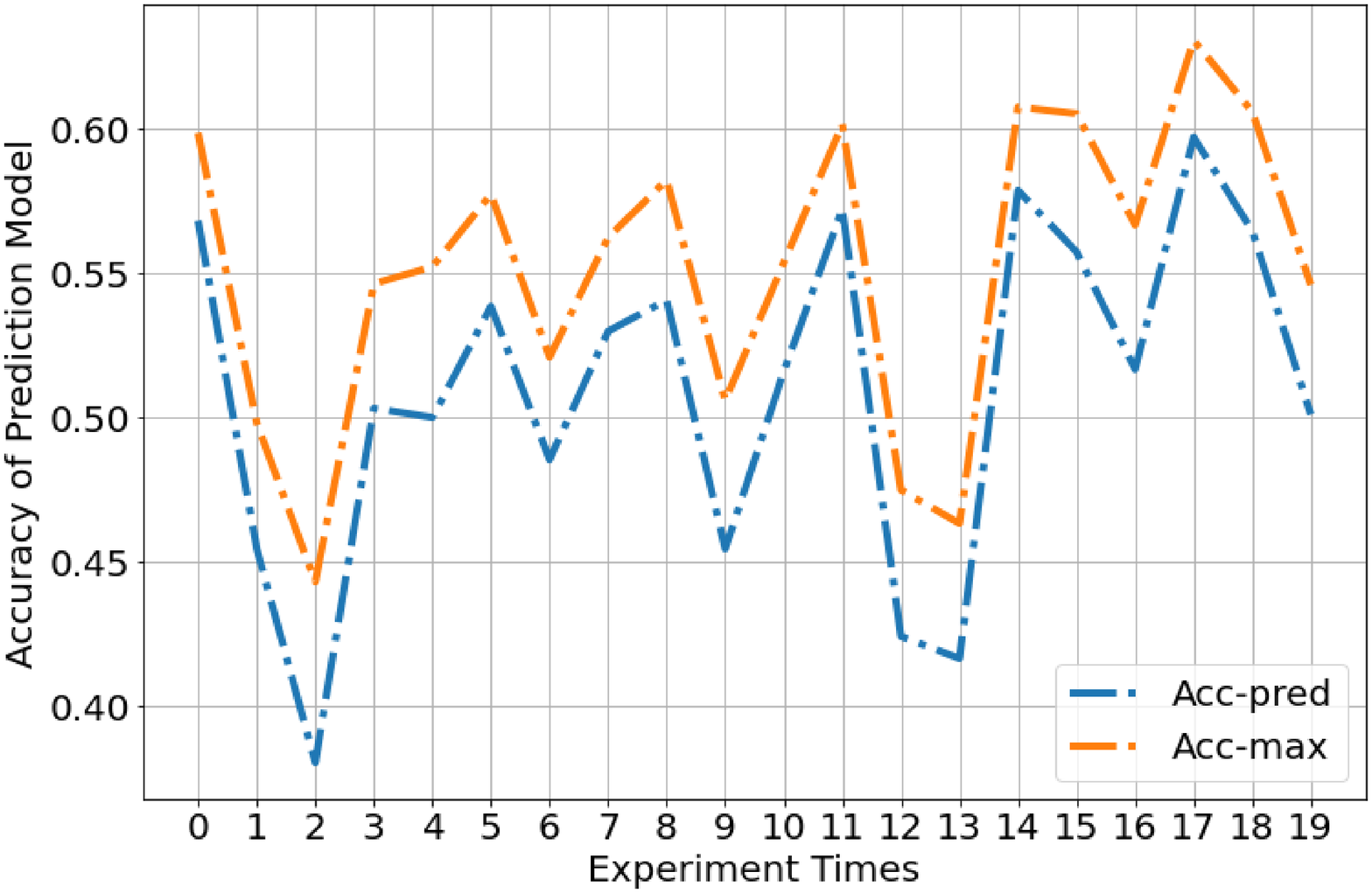}
\centering{\caption{The accuracy of LQ prediction model with randomness selected training sets:Four-classification case}\label{}}
\end{figure}

\begin{figure}[h]
\centering
\vspace{0.1cm}
\includegraphics[width=3.2 in]{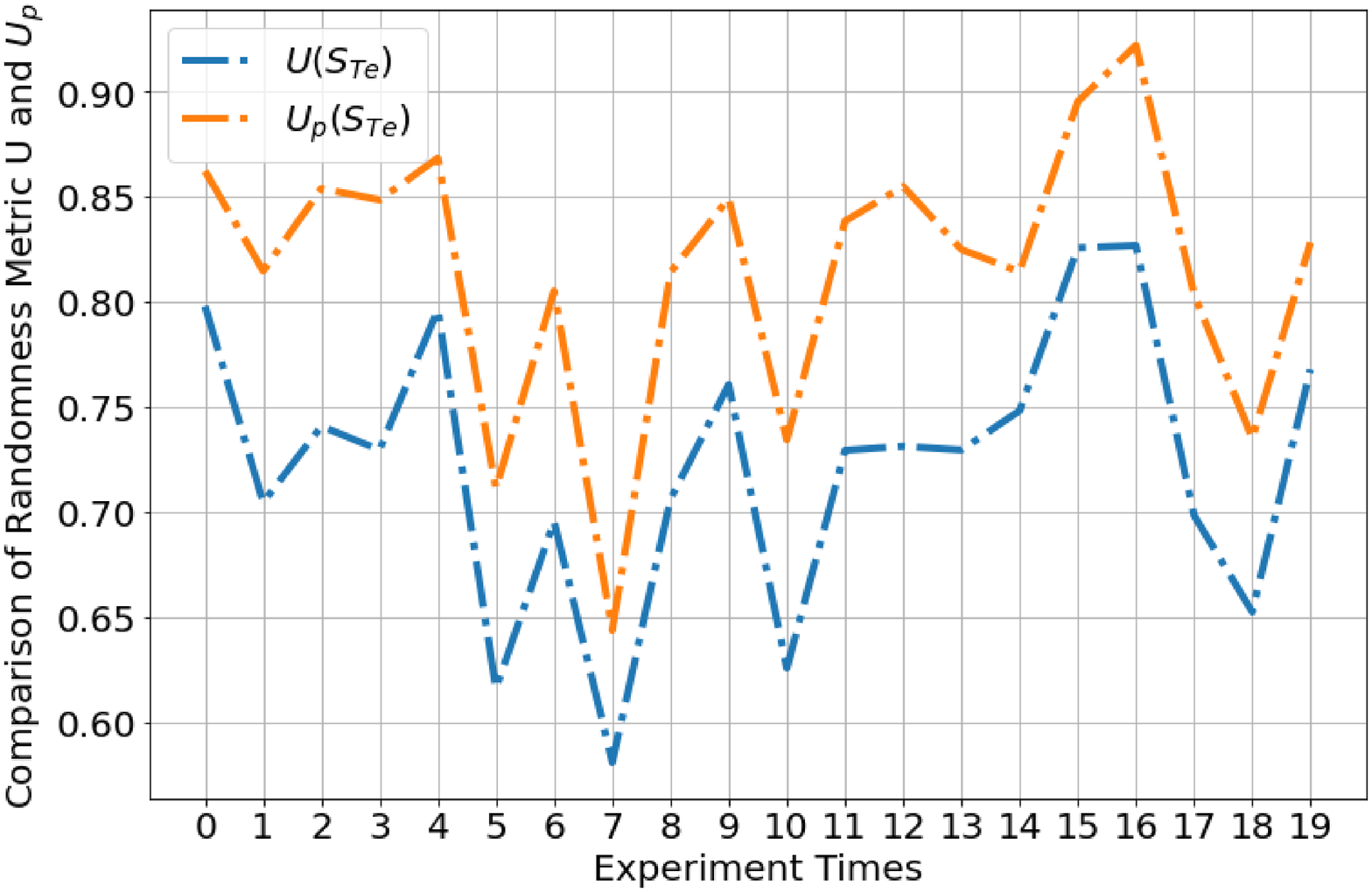}
\centering{\caption{The randomness metric of LQ prediction model with randomness selected training sets:Two-classification case}\label{}}
\end{figure}

\begin{figure}[h]
\centering
\vspace{0.1cm}
\includegraphics[width=3.2 in]{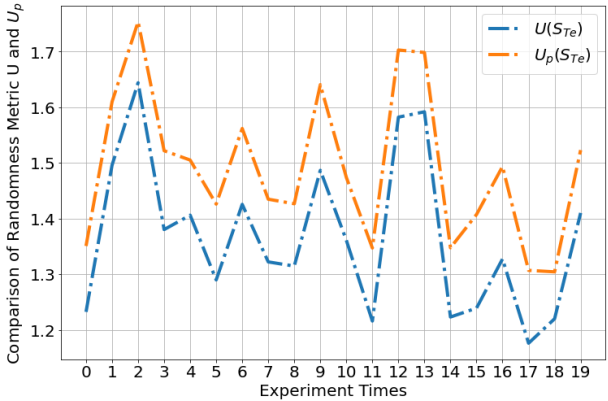}
\centering{\caption{The randomness metric of LQ prediction model with randomness selected training sets:Four-classification case}\label{}}
\end{figure}

Fig 12 and Fig 13 are the comparison of the accuracy of prediction model and the reference maximum accuracy for two-classification and four-classification cases. In each experiment, the accuracy of the prediction model is slightly less than the reference maximum accuracy. The corresponding randomness metric $U$ and $U_p$ is as shown in Fig 14 and Fig 15. From the figures, we know that the randomness of prediction output is always stronger than the randomness of input sample set, which means that the predictor brings additional randomness and the performance is limited. 

In conclusion, since the randomness of the wireless transmission cannot be eliminated, the randomness of the prediction output cannot be eliminated either. The accuracy of the LQ prediction model is restricted by the randomness of the wireless link, and it is impossible to achieve a completely accurate prediction, especially at the edge of wireless coverage with strong randomness, the output of LQ prediction model becomes purely random.

\section{Application of LQ prediction model }
Although the prediction model is not completely accurate to predict the link situation at the next moment, we can still use the prediction output as the basis for whether the node processes the received data packets, which helps greatly to improve the stability of wireless transmission. The two-classification case for example, when the output of the prediction model is received, and the corresponding data packet is received at the next moment, the data packet will be processed. Otherwise, if the output is lost, even if the data packet is received at the next moment, it will be discarded.

\begin{figure}[h]
\centering
\vspace{0.1cm}
\includegraphics[width=3.2 in]{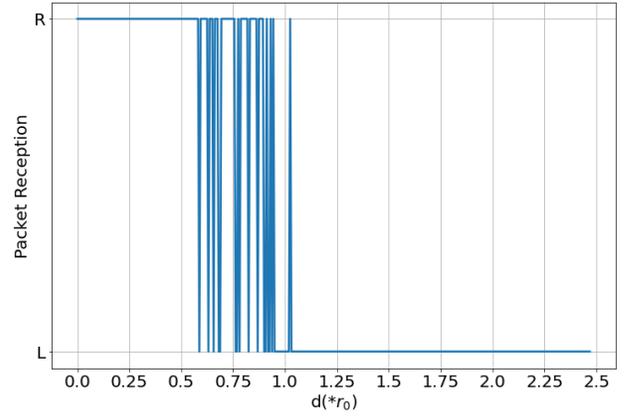}
\centering{\caption{The packet reception of a node with LQ prediction model}\label{}}
\end{figure}

We train a two-classification LQ prediction model and deploy on the nodes. The packet reception of the node with LQ predoction model is as shown in Fig 16. From the figure, we can see that the packet reception still has a certain randomness. However, compare with Fig 5, the randomness is weaker. Moreover, the strongest randomness occurs at the position where the distance is $0.75r_0$. This is because in an area with strong randomness, when the prediction output is lost, the packet will be discarded even it is successfully received. Therefore, the packet delivery rate is further decrease. Actually, the prediction model helps to discard the accidentally received packets from the weak connections. The same law is also reflected in the curves of packet delivery rate in the cases before and after using prediction model. 

\begin{figure}[h]
\centering
\vspace{0.1cm}
\includegraphics[width=3.2 in]{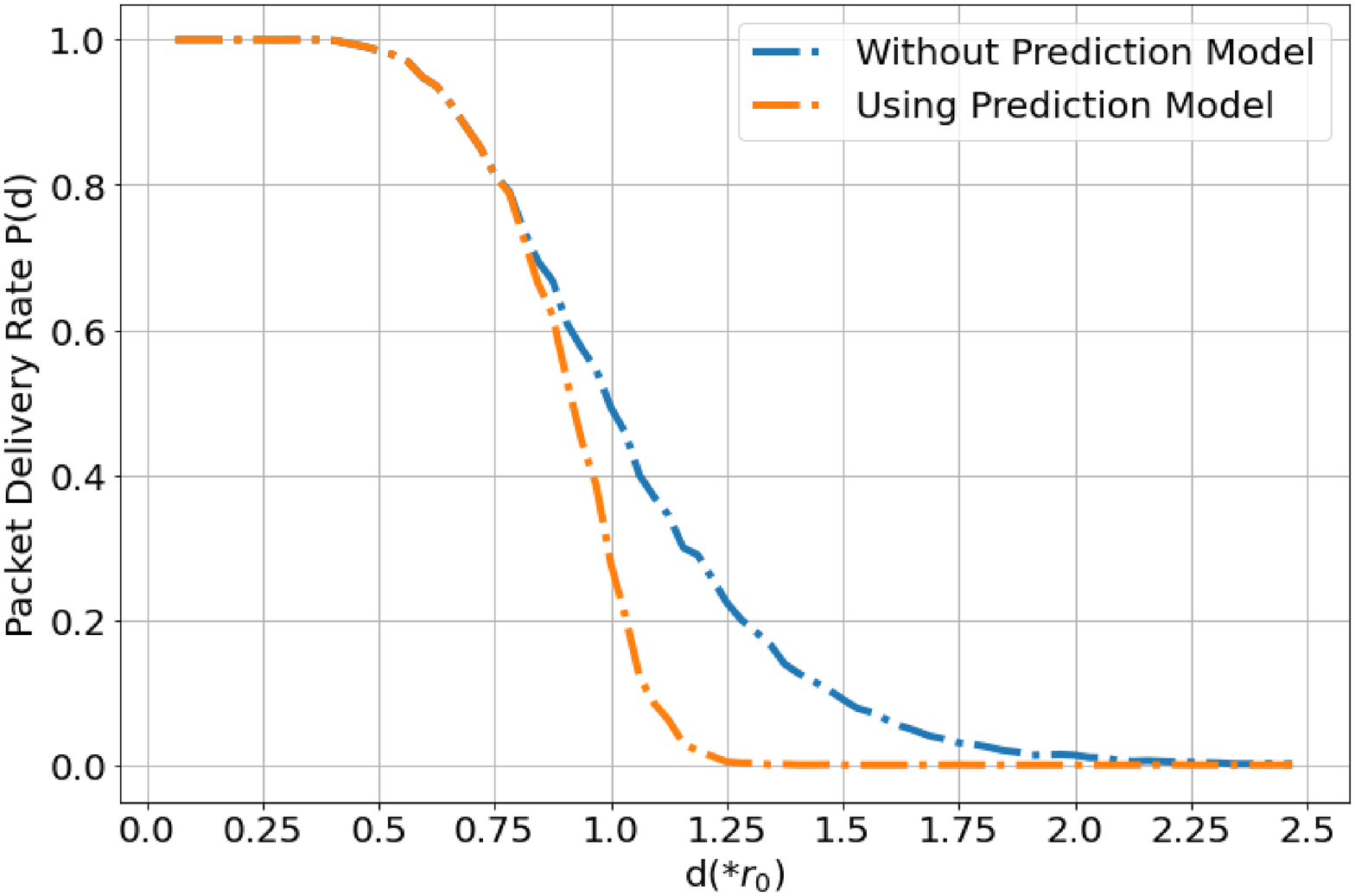}
\centering{\caption{The comparison of packet delivery rate before and after using prediction model}\label{}}
\end{figure}

Fig 17 is the comparison of packet delivery rate before and after using prediction model. After using the prediction model, the packet delivery rate drops sharply, which means that the area with strong randomness is greatly reduced. In wireless multi-hop networks, this helps to reduce the use of unstable links and improve the network stability. 

\begin{figure}[h]
\centering
\vspace{0.1cm}
\includegraphics[width=3.2 in]{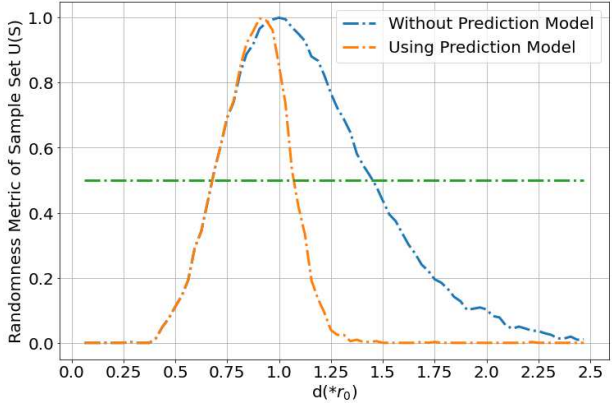}
\centering{\caption{The comparison of randomness metric before and after using prediction model}\label{}}
\end{figure}

The comparison of randomness metric $U$ before and after using prediction model is as shown in Fig 18. If we set the threshold of $U$ as $U_{th}=0.5$, it is easy to see that after using the prediction model, the area with strong randomness is greatly reduced. 

Fig 19-21 are the packet receptions before and after using prediction model at nodes with distance $0.8r_0$,$r_0$ and $1.2r_0$ respectively. It is intuitively shown that after using prediction model, the packet reception becomes more stable in spite of the low prediction accuracy. When $d=0.8r_0$, the prediction outputs are almost received, which means almost all received packets will be handed over to the upper-layer. When $d=1.2r_0$, the prediction outputs are almost lost. As a result, the receiver discards accidentally received packets. 

\begin{figure}[h]
\centering
\vspace{0.1cm}
\includegraphics[width=3.2 in]{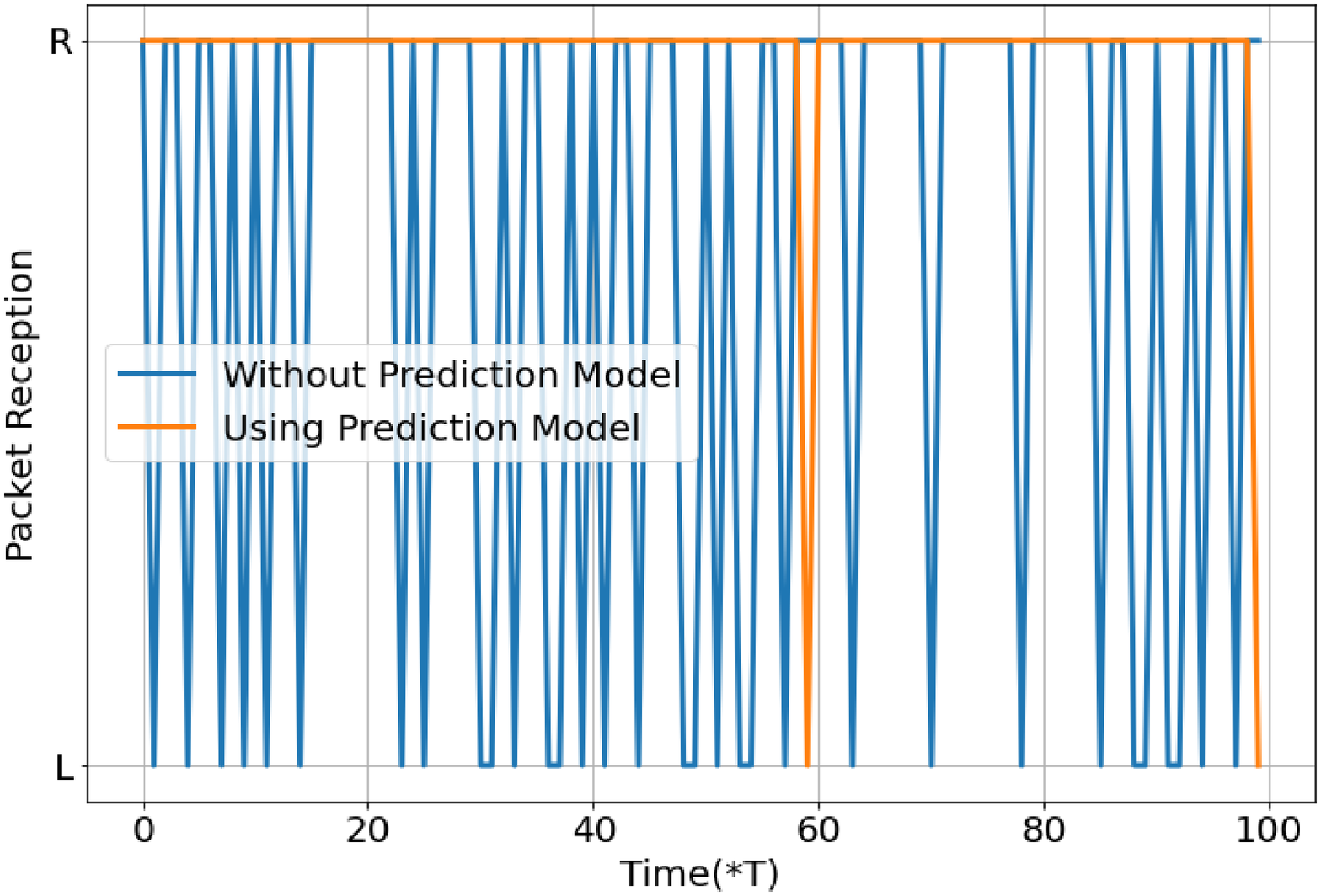}
\centering{\caption{The packet reception before and after using prediction model:$d=0.8r_0$}\label{}}
\end{figure}

\begin{figure}[h]
\centering
\vspace{0.1cm}
\includegraphics[width=3.2 in]{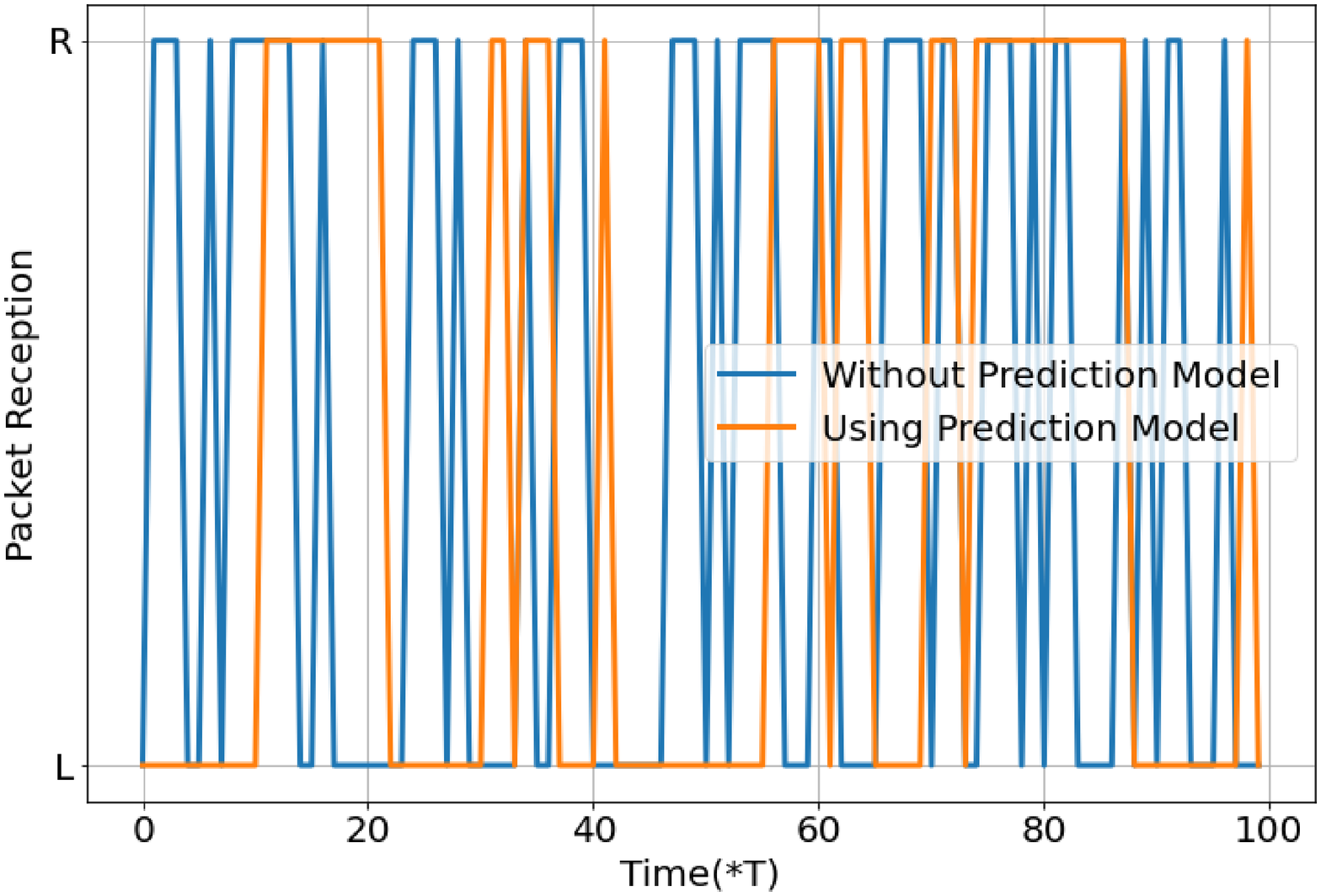}
\centering{\caption{The packet reception before and after using prediction model:$d=r_0$}\label{}}
\end{figure}

\begin{figure}[h]
\centering
\vspace{0.1cm}
\includegraphics[width=3.2 in]{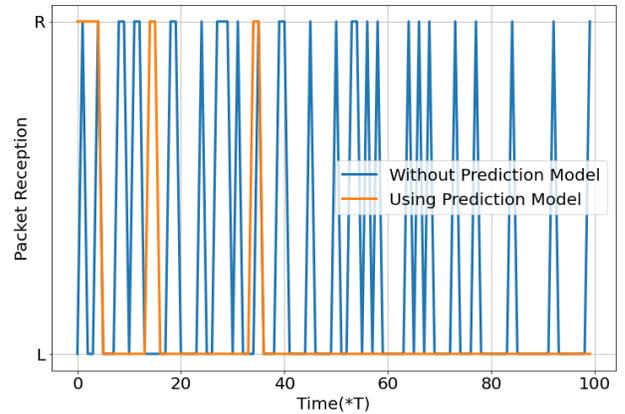}
\centering{\caption{The packet reception before and after using prediction model:$d=1.2r_0$}\label{}}
\end{figure}

\section{Conclusion}
In this paper, we mainly analyze the impact of randomness of wireless transmission on the LQ prediction models. Firstly, we propose a two-classification and a four-classification LQ prediction models. The proposed models apply the combination of the physical and logical metrics as input parameters. Secondly, we defined the randomness metric of both the sample set and the prediction model. Then we discuss the reasonableness of the definitions. At last, we analyze the performance of the proposed models under environments with different randomness and discuss the benefits of the LQ prediction model.

Our research shows that, the randomness of wireless transmission has great impact on the performance of LQ prediction models and the accuracy of the prediction model is limited. In areas with strong randomness, the link prediction output is even a purely random process.  However, in the application of LQ prediction, the use of the model still has positive effects and helps to improve the stability of wireless multi-hop networks.

\section*{Conflict of Interests}
The authors declare that there is no conflict of interest regarding the publication of this article.

\section*{Acknowledgment}

This work is supported by Information Engineering College Shaoguan University.



\end{document}